\documentstyle[preprint,aps]{revtex}
\input psfig
\newcommand{\psplot}[1]{ \hbox to\textwidth{\hfill
             \psfig{figure=#1.ps,width=4.0in,height=4.0in}\hfill} }
\begin{document}

\draft

\tighten 
\preprint{\vbox{\hbox{CLNS 94/1267 \hfill}  
               \hbox{CLEO 94-2     \hfill}
               \hbox{\today        \hfill}}}

\title{Measurement of the branching fraction for
$D^+ \rightarrow K^-\pi^+\pi^+$}

\author{
R.~Balest,$^{1}$ K.~Cho,$^{1}$ M.~Daoudi,$^{1}$ W.T.~Ford,$^{1}$
D.R.~Johnson,$^{1}$ K.~Lingel,$^{1}$ M.~Lohner,$^{1}$ P.~Rankin,$^{1}$
J.G.~Smith,$^{1}$
J.P.~Alexander,$^{2}$
% ...
}

\address{
\bigskip 
{\rm (CLEO Collaboration)}\\  
\newpage 
$^{1}${University of Colorado, Boulder, Colorado 80309-0390}\\
$^{2}${Cornell University, Ithaca, New York 14853}\\
% ...
\bigskip 
}        

\date{\today}
\maketitle


\begin{abstract}
Using the CLEO-II detector at CESR we have measured the ratio of branching
fractions,
${\cal B}(D^+\rightarrow K^- \pi^+ \pi^+)/{\cal B}(D^0 \rightarrow K^-\pi^+)
= 2.35 \pm 0.16 \pm 0.16$.
Our recent measurement of
${\cal B}(D^0 \rightarrow K^-\pi^+)$ then gives
${\cal B}(D^+\rightarrow K^- \pi^+ \pi^+) = (9.3 \pm 0.6 \pm 0.8)\%$.
\end{abstract}


   The decay $D^+ \rightarrow K^- \pi^+ \pi^+$ is the most commonly
used mode for normalizing $D^+$ yields, since it has a relatively large
branching fraction and is one of the simplest to reconstruct.
Many current charm and
bottom meson decay results are limited by the precision of
${\cal B}(D^+ \rightarrow  K^- \pi^+ \pi^+)$.
Previous measurements of this decay mode were performed by the Mark III
\cite{mark} and ACCMOR \cite{accm} collaborations.
Mark III used the relative number of singly detected $D^{\pm}$ mesons to the
number of reconstructed $D^+ D^-$ events to determine the branching fraction.
The
ACCMOR collaboration measured the ratio of $D^+ \rightarrow K^- \pi^+ \pi^+$
relative to the total number of 3-prong decays, and used
topological branching ratios determined by other experiments to obtain a
branching fraction. However, ACCMOR could not easily distinguish $D^+, D_s^+$
and
$\Lambda_c^+$ decay vertices, and had to rely on
estimates of the relative production ratios of these particles. In this
analysis,
we use the exclusive yields, $N_{K\pi}$ and $N_{K\pi\pi}$, of the
$(D^{*+} \rightarrow D^0 \pi^+, D^0 \rightarrow K^- \pi^+)$ and
the $(D^{*+} \rightarrow D^+ \pi^0, D^+ \rightarrow K^- \pi^+ \pi^+)$ decay
sequences, respectively, to measure the ratio,
${\cal B}(D^+ \rightarrow K^- \pi^+ \pi^+)/{\cal B}(D^0 \rightarrow K^-
\pi^+)$,
and apply our measurement of the branching fraction for
${\cal B}(D^0 \rightarrow K^- \pi^+)$ \cite{kpi} to obtain
${\cal B}(D^+ \rightarrow K^- \pi^+ \pi^+)$.


   The data used in this analysis consist of 1.79 fb$^{-1}$ of $e^+e^-$
collisions recorded with the CLEO-II detector operating at the Cornell Electron
Storage Ring (CESR). The CLEO-II detector has been described in detail
elsewhere
\cite{nim}. Data were recorded at the $\Upsilon(4S)$ resonance and in the
continuum both below and above (the $e^+e^-$ center of mass energies ranged
from
10.52 to 10.70 GeV).

  We obtain clean samples of $D^*$ mesons by requiring the $\pi^0$
and the $\pi^+$ emitted in their decays to fulfill strict selection criteria.
To reconstruct $\pi^0$'s, we start with neutral showers which satisfy
isolation cuts and cannot be matched to any charged track in the event.
These photons candidates must have $|\cos \theta_{\gamma}| \le 0.71$,
($\theta_{\gamma}$ is the polar angle measured relative to the beam axis) to
ensure
that they lie in that portion of the electromagentic calorimeter which has the
best
efficiency and resolution, and the least systematic uncertainty.
In addition, photon energies have to be greater than 30 MeV. We then
kinematically
constrain $\gamma\gamma$ combinations with masses between
125 and 145 MeV/c$^2$ to the known $\pi^0$ mass to improve the momentum
resolution. To reduce $\gamma\gamma$ combinatoric background,
$\pi^0$ candidates are required to have momenta greater than 200 MeV/c. In
addition, the kinematically constrained $\pi^0$ candidates must
have $|\cos \theta_{\pi^0}| \le 0.70$. Charged pions
are selected if they have momentum greater than 200 MeV/c, and
$|\cos \theta_{\pi^{\pm}}| \le 0.70$. The polar angle cuts on $\pi^+$'s and
$\pi^0$'s
ensure that $D^{*+}$
mesons reconstructed with either charged or neutral pions have the
same geometric acceptance.


The ratio $N_{K\pi\pi}/N_{K\pi}$ of the measured yields can be expressed in
terms of
branching ratios and efficiencies as,
\begin{equation}
\frac
{N_{K\pi\pi}}
{N_{K\pi}} =
\frac
{N_{D^{*+}}{\cal B}(D^{*+}\rightarrow D^+\pi^0) {\cal
B}_{K\pi\pi}\epsilon_{K\pi\pi}}
{N_{D^{*+}}{\cal B}(D^{*+}\rightarrow D^0\pi^+) {\cal B}_{K\pi}\epsilon_{K\pi}}
\label{eq:one}
\end{equation}

\noindent where ${\cal B}_{K\pi\pi}$ and ${\cal B}_{K\pi}$ are the relevant
$D^+$ and
$D^0$ branching fractions, respectively.
The total number of $D^{*+}$'s produced in the data sample is $N_{D^{*+}}$
(which
cancels in the ratio); $\epsilon_{K\pi\pi}$ and $\epsilon_{K\pi}$ are the
efficiencies
for reconstructing $D^+ \rightarrow K^- \pi^+ \pi^+$ and $D^0\rightarrow K^-
\pi^+$,
respectively, with their respective $D^{*+}$ tags. Using isospin invariance,
the
CLEO-II measurements of the $D^{*+} - D^+$ and $D^{*+} - D^0$ mass
differences \cite{massd}, and the fact that these decays are p-wave, we
estimate the ratio \cite{pwave}
\begin{equation}
\frac
{{\cal B}(D^{*+}\rightarrow \pi^+D^0)}
{{\cal B}(D^{*+} \rightarrow \pi^0 D^+)} = 2.21 \pm 0.07
\label{eq:two}
\end{equation}
\noindent The efficiencies in Eq.\ (\ref{eq:one}) include the
efficiency of reconstructing the $D$ decay, as well as the efficiency for the
slow pion emitted in the
$D^*$ decay. The $D$ reconstruction efficiency is reliably simulated by the
Monte
Carlo because the main cuts on the $D$ daughters are
geometric. It is harder to simulate the efficiencies to detect the slow neutral
and
charged pions, since the efficiency for detecting charged tracks varies rapidly
at low
momentum \cite{mom}, and the $\pi^0$ efficiency is known only to $\pm 5\%$.
We have checked the slow charged and neutral pion efficiencies from the data in
several ways. The ratio of branching fractions for
$\eta \rightarrow \pi^0 \pi^0 \pi^0$ and $\eta \rightarrow \gamma \gamma$ as
measured in our data sample has been
compared with the world average \cite{pdg} to obtain an estimate of the
accuracy of
the photon finding efficiency.  We find that this efficiency is simulated to
an accuracy of $\pm 2.5\%$.
We have studied the charged
particle tracking by comparing the yield of fully reconstructed
$D^0 \rightarrow K^- \pi^+ \pi^0$ with partially reconstructed
$D^0 \rightarrow K^- \pi^0 (\pi^+)$, where the $\pi^+$ is not detected. This
check
shows that the charged pion efficiency is simulated to a precision which is
better
than $\pm 2\%$.

   In addition, we have directly checked the ratio of slow neutral and charged
pion
efficiencies, $\epsilon_{\pi^0}/\epsilon_{\pi^+}$, from the data.
 Using yields for $D^{*+(0)} \rightarrow D^0 \pi^{+(0)}$,
where $D^0 \rightarrow K^- \pi^+$, we have
measured the ratio of inclusive $D^{*+}$ and $D^{*0}$ production cross-section
in
the continuum to be $1.06 \pm 0.09$. This result is consistent with unity,
which is
expected since
the $D^*$'s are not being produced near threshold and the $D^{*+} - D^{*0}$
mass difference is negligible compared to the center of mass energy.
 We have also
studied this ratio by using the decays
$\eta \rightarrow \pi^+ \pi^- \pi^0$ and $K^0_s \rightarrow \pi^+\pi^-,
\pi^0\pi^0$;
here we find that the Monte Carlo estimate for
$\epsilon_{\pi^0}/\epsilon_{\pi^+}$
is $0.96\pm 0.05$ times the value extracted from data. In summary, all our
checks of the estimate for $\epsilon_{\pi^0}/\epsilon_{\pi^+}$
show that there is good agreement between Monte Carlo and data.
We assign a systematic error of $\pm 5.4\%$ due to this efficiency ratio in the
final
results.

   We now determine the exclusive yields and reconstruction efficiencies
used in Eq.\ (\ref{eq:one}).
   The $D^+$ is reconstructed by requiring the $ K^-$ and $ \pi^+$ tracks
to have $|\cos\theta| \le 0.81$, and
to have momenta greater than 200 MeV/c. In Fig. ~\ref{fig:dspdp}
we present the invariant mass distribution of $ K^- \pi^+ \pi^+$
combinations, for which the mass difference,
$\Delta_{M^0} = M_{K^- \pi^+ \pi^+(\pi^0)} - M_{ K^- \pi^+ \pi^+}$,
is within $\pm 5$ MeV/$c^2$ of the value expected for $M_{D^{*+}} - M_{D^+}$.
Using two Gaussians for the signal and a first order polynomial for the
background, we obtain $1618 \pm 91$ events.
To account for true $D^+$ - random $\pi^0$ combinations, we also fit the
$\Delta_{M^0}$
sidebands. We observe $116 \pm 40$ events in the scaled sideband \cite{scale}.
After subtraction, the net yield is $1502 \pm 99$ events.
The efficiency $\epsilon_{K\pi\pi}$ is estimated from Monte Carlo to be
$(15.4 \pm 0.2) \%$.

   The $D^0$ is tagged using the
$D^{*+} \rightarrow D^0 \pi^+$ decay, and is reconstructed by
requiring the $ K^-$ and $ \pi^+$ tracks
to have $|\cos\theta| \le 0.81$, and to have momenta greater than 200 MeV/c.
In Fig. ~\ref{fig:dspdz} we present the invariant mass distribution of $ K^-
\pi^+$
combinations, where the mass difference,
$\Delta_{M^+} = M_{K^- \pi^+ (\pi^+)} - M_{ K^- \pi^+}$,
is within $\pm 5$ MeV/$c^2$ of the value expected for $M_{D^{*+}} - M_{D^0}$.
Using two Gaussians for the signal and a first order polynomial for the
background, we obtain $5555 \pm 102$ events.
A fit to the $\Delta_{M^+}$ sidebands yields $103 \pm 21$ events \cite{scale}.
The net yield is $5452 \pm 104$ events.
The efficiency $\epsilon_{K\pi}$ is estimated to be $(59.5 \pm 0.6) \%$.

    Substituting the exclusive yields, the efficiencies for
reconstructing these final states, and the known ratio of $D^{*+} \rightarrow
D\pi$
branching fractions in Eq.\ (\ref{eq:one}), we obtain
\begin{equation}
\frac
{{\cal B}(D^+ \rightarrow K^- \pi^+ \pi^+)}
{{\cal B}(D^0 \rightarrow K^- \pi^+)} = 2.35 \pm 0.16 \pm 0.16,
\label{eq:three}
\end{equation}
\noindent where the first error is statistical and the second is an estimate of
the
systematic
uncertainty. The systematic error includes the uncertainty on the ratio of
$D^{*+} \rightarrow D\pi$ branching fractions ($\pm 3.2\%$), the error due to
$\epsilon_{\pi^0}/\epsilon_{\pi^+}$ ($\pm 5.4\%$), the
error due to Monte Carlo statistics ($\pm 1.6\%$), and the uncertainty
from the effects of resonant substructure on the $K\pi\pi$ final state ($\pm
1.3\%$).

\noindent Using our measurement \cite{kpi}, ${\cal B}(D^0 \rightarrow K^-
\pi^+) =
(3.95 \pm 0.08 \pm 0.17)\%$, we obtain
\begin{equation}
{\cal B}(D^+ \rightarrow K^- \pi^+ \pi^+) = (9.3 \pm 0.6 \pm 0.8)\%
\label{eq:four}
\end{equation}
\noindent This result accounts for the effects of decay radiation in the final
state, because we have used the radiatively corrected value for
${\cal B}(D^0 \rightarrow K^- \pi^+)$ \cite{decr}.
The determination of statistical and systematic errors is described above. The
systematic error also includes the error in our measurement of
${\cal B}(D^0 \rightarrow K^- \pi^+)$.


    In conclusion, using yields of $(D^+ \rightarrow K^-\pi^+\pi^+)$ and
$(D^0 \rightarrow
K^-\pi^+)$, which have been tagged via $D^{*+} \rightarrow D\pi$ decays,
and our measurement of ${\cal B}(D^0 \rightarrow  K^- \pi^+)$ \cite{kpi}, we
obtain ${\cal B}(D^+ \rightarrow  K^- \pi^+ \pi^+) = (9.3 \pm 0.6 \pm 0.8) \%$.
   This result agrees well with the Mark III measurement,
$(9.1\pm1.3 \pm 0.4)\%$ \cite{mark}, but is larger than the ACCMOR result,
$(6.4 \pm 1.5)\%$ \cite{accm}.

\smallskip
We gratefully acknowledge the effort of the CESR staff in providing us with
excellent luminosity and running conditions.
This work was supported by the National Science Foundation, the U.S. Dept. of
Energy, the SSC Fellowship of TNRLC, the Heisenberg Foundation and the A.P.
Sloan Foundation.


%
%
%
%
%

\begin{figure}
\caption{Mass distribution for $D^+ \rightarrow  K^- \pi^+ \pi^+$ candidates
tagged via $D^{*+} \rightarrow D^+ \pi^0$ decays;
the histogram represent events in the mass difference signal region, triangles
with
error bars represent events in the (scaled) mass difference sideband region.
The solid line is the fit to the data.}
\label{fig:dspdp}
\end{figure}
%
\begin{figure}
\caption{Mass distribution for $D^0 \rightarrow  K^- \pi^+$
tagged via $D^{*+} \rightarrow D^0 \pi^+$ decays;
the histogram represent events in the mass difference signal region, triangles
represent events in the (scaled) mass difference sideband region.
The solid line is the fit to the data.}
\label{fig:dspdz}
\end{figure}
%


\begin{references}
\bibitem{mom} The average momentum for slow pions emitted in $D^*$ decays which
pass the selection criteria is approximately 260 MeV/c, whereas the average
momenta of the $D$ decay daughters is approximately 1.3 GeV/c for $D^+$ and
1.7 GeV/c for
$D^0$. The efficiency for reconstructing kaons and pions from $D$ decays is
known to an
accuracy of $\pm 2\%$ \cite{kpi}.
\bibitem{scale} The scale factor is such that the number of events in the
sideband region is the same as the number of events in the background under the
mass difference signal. We investigated different mass difference sidebands
to estimate the systematic error associated with this technique. The yields
quoted in
the text are obtained from the sideband where the $D^{*+} - D$ mass difference
is in the
range 146.6 - 155.6 MeV/$c^2$ for $D^{*+} - D^+$, and 151.4 - 160.4 MeV/$c^2$
for
$D^{*+} - D^0$.
\end{references}
\end{document}